\documentclass[twocolumn,aps,prd,10pt]{revtex4-2}

\usepackage[dvipsnames]{xcolor}
\usepackage{amsmath,amssymb,bm}
\usepackage{graphicx}
\usepackage{hyperref}
\hypersetup{
    colorlinks = true,
    linkcolor = blue,
    anchorcolor = blue,
    citecolor = blue,
    filecolor = blue,
    urlcolor = blue
    }
\usepackage{float}
\usepackage{subcaption}
\usepackage{caption}
\usepackage{braket}
\usepackage{booktabs}
\usepackage[normalem]{ulem}

\begin{document}

\title{Temporal processing of quantum states with hybrid quantum–classical reservoirs}
\author{Mateu Coll-Comas}
\email{mateu@ifisc.uib-csic.es}
\author{Gian Luca Giorgi}
\author{Roberta Zambrini}
\affiliation{Institute for Cross-Disciplinary Physics and Complex Systems (IFISC) UIB-CSIC \\
Campus Universitat Illes Balears, 07122, Palma de Mallorca, Spain.}
\date{\today}

\begin{abstract}
A distinctive feature of Quantum Reservoir Computing (QRC) is the ability to directly embed quantum input states into the reservoir dynamics. However, the resulting output is fundamentally linear for a single input state, preventing QRC from naturally computing nonlinear functionals such as purity or entropy. We overcome this limitation with a quantum-classical hybrid architecture combining a qubit quantum reservoir with a classical echo state network (ESN), allowing both nonlinear functional approximation and effective temporal processing. We systematically study performance under two information regimes: full-tomography and partial information (single-axis measurements), with the latter demonstrating that the hybrid system outperforms its standalone components in both linear and nonlinear tasks due to the enhanced information retrieval provided by the quantum reservoir. Building on these results, we apply an \textit{online monitoring protocol} that explicitly accounts for measurement back-action and finite measurement ensembles, enabling a realistic assessment of performance under experimental conditions. These results establish hybrid quantum–classical reservoir computing (HRC) architectures as a practical and scalable route for enhanced quantum machine learning on near-term qubit hardware.
\end{abstract}

\maketitle

\section{Introduction}

Reservoir computing (RC) is a supervised machine learning paradigm in which only the output layer rooted in recurrent neural networks is trained, while the internal dynamics of the reservoir remain fixed \cite{Jaeger2001TheechoST,Maass2002RealTimeCW}. This approach offers stable training, low computational cost, and strong performance in temporal processing tasks \cite{Luko_2009, Van_Der_Sande2017Advances}. Importantly, the RC framework allows a wide range of dynamical systems to act as reservoirs \cite{Appeltant2011, Hauser2012TheRO, Vandoorne2014, Fujii2016, goto2020computingvorticesbridgingfluid, Nakajima_2020, Chen2023All-ferroelectric}, provided they satisfy the echo state property, separability, and fading memory \cite{adamatzky2016advances, Grigoryeva2018Echo}.
This physical flexibility has motivated  the development of quantum reservoir computing (QRC) \cite{Mujal_2021, Fujii2021}, where quantum systems are employed as reservoirs \cite{Fujii2016, Govia2021,  Nokkala2021, Spagnolo2022, PRXQuantum.3.030325, Garc_a_Beni_2023, Hu2024, kornjača2024largescalequantumreservoirlearning, Senanian2024, das2025quantumreservoircomputingusing, LLODRA2025116289, carles2026experimentalquantumreservoircomputing, Paparelle2026}. Quantum reservoirs exploit the exponential growth of Hilbert space with system size, enabling high-dimensional feature representations through both accessible and hidden degrees of freedom.

While QRC has been demonstrated to have promising performance for classical input time series \cite{ Mujal_2021, Nokkala2021, Suzuki2022, Čindrak2023Enhancing, Mujal2023, Dudas2023, PhysRevA.108.052427, Zhu2024Practical, Sannia2024dissipationas, Martinez2025}, its extension to quantum inputs \cite{Ghosh2019, 9153954, Ghosh_2021, Ghosh2021, PhysRevLett.127.260401, Nokkala2023, Zia2025} exhibits specific challenges. A key advantage of QRC is that quantum input states can be directly embedded into the reservoir dynamics, avoiding the need for full state tomography\textemdash a feature that has no classical counterpart. However, the dependence of a quantum reservoir on its input state is fundamentally linear. As a consequence, while QRC is well-suited for tasks involving linear functionals of the input density matrix \cite{Innocenti2023}, it cannot natively perform nonlinear quantum tasks such as purity or entropy estimation. This contrasts with the classical-input setting, where nonlinear encoding and processing strategies are possible \cite{Govia2021Nonlinear, Mujal_2021_Analytical}.
This restriction is significant because many physically relevant properties of quantum states—such as purity, entropy, or entanglement witnesses—are nonlinear functionals of the density matrix. Achieving nonlinear processing of quantum inputs while retaining temporal memory, therefore, remains a key challenge.

Recent works have considered quantum extreme learning machines (QELMs) and have shown that nonlinear functionals of static quantum inputs can be learned via multiple-injection or distributed architectures \cite{Innocenti2023, gili2026learningfunctionsquantumstates}. However, these algorithms are inherently limited to static tasks and cannot handle temporal tasks requiring memory.

Achieving nonlinear processing of quantum inputs in a dynamical (temporal) setting requires architectures beyond purely quantum reservoirs. In particular, this motivates the exploration of hybrid quantum–classical approaches. In the hybrid architecture proposed in \cite{Nokkala2024}, a photonic quantum reservoir with Gaussian states is used as a preprocessor that transforms quantum inputs into classical representations (obtained from the covariance matrix), which are then fed into the classical echo state network (ESN) \cite{Jaeger2001TheechoST} for further processing. While this work provided an initial demonstration of the hybrid concept for processing quantum inputs, it opens several questions about the potential advantage of the hybrid approach beyond Gaussian photonic settings.

In contrast to \cite{Nokkala2024}, where the quantum reservoir operates in an infinite-dimensional Hilbert space, our approach focuses on finite-dimensional quantum systems\textemdash specifically qubit-based systems\textemdash allowing us to investigate hybrid quantum-classical reservoir computing (HRC) in experimentally relevant QRC settings. We will show that in this configuration, the quantum reservoir distributes information across its degrees of freedom without requiring full state tomography, while the classical reservoir provides strong nonlinear processing and long-term memory. Operating the quantum reservoir in the ergodic regime, which has been shown to be advantageous for QRC \cite{Pena2021}, enables efficient feature extraction while avoiding direct measurements of non-commuting observables.

The impact of noise, dissipation, and measurement back-action on reservoir performance needs to be studied to establish the operational value of the HRC to process quantum (and even classical) inputs. Previous studies have shown that dissipation or decoherence can enhance QRC performance under specific conditions \cite{PhysRevApplied.14.024065, Suzuki2022, PhysRevResearch.5.023057, Domingo2023, Sannia2024dissipationas, Mifune2024Effects, Cheamsawat2024Dissipation, Palacios_2024}. In realistic online protocols, a strategy based on weak measurements was put forward in \cite{Mujal2023}, also showing the effect of statistical noise arising from finite measurement trajectories.

We notice that in the hybrid approach we propose, quantum and classical machine learning layers complement each other. In other designs, hybrid refer to the combination of classical and quantum operations in the full dynamical map, generally targeting classical input processing \cite{Pfeffer_2022, Pfeffer2023Reduced-order, wudarski2024hybridquantumclassicalreservoircomputing, Kobayashi_2024, Sakurai_2025, Settino_2025, monomi2025feedbackenhancedquantumreservoircomputing, Paparelle2026}.

In this work we propose an operative HRC qubits architecture for non-linear temporal tasks of quantum inputs. In section~\ref{sec:Model}, we introduce the hybrid model and describe its individual components and workflow. Sections~\ref{sec: Information Spreading QRC} and~\ref{sec:Hybrid RC Performance} analyze the performance of the HRC, focusing on the measurement-direction dependence of information retrieval in the quantum reservoir and evaluating the full system’s performance on both nonlinear and linear tasks. We also investigate the scaling of these results for a two-qubit input. In section~\ref{sec:State Perturbation Effects}, we consider a more realistic protocol that incorporates back-action and statistical uncertainty arising from weak measurements and the finite number of trajectories, in order to assess their impact on performance. Finally, section~\ref{sec:Conclusions} summarizes our findings and outlines possible directions for future research.

\section{Hybrid quantum-classical reservoir computing}
\label{sec:Model}

In this section, we present the HRC architecture, describing the quantum and classical reservoir computing layers individually and then outlining how they are integrated into a unified workflow.

\subsection{QRC layer}

The quantum reservoir employed in this work is defined on an Ising Hamiltonian network of $L$ spins with a transverse field $h$ and site-dependent local disorder $D_i$:
\begin{equation}
    \label{eq:Ising_Hamiltonian}
    H = \sum_{i>j=1}^L J_{ij}\,\sigma^x_i \sigma^x_j 
        + \frac{1}{2}\sum_{i=1}^L (h + D_i)\,\sigma^z_i ,
\end{equation}
where $\sigma_i^\alpha$ denotes the Pauli matrix along the axis $\alpha \in \{x,y,z\}$ acting on spin $i$, i.e.\ 
$\sigma_i^\alpha = \mathbb{I}_1 \otimes \cdots \otimes \sigma_i^\alpha \otimes \cdots \otimes \mathbb{I}_L$. 
The disorder terms $D_i$ are drawn from a uniform distribution $[-W, W]$, and the couplings $J_{ij}$ from $[-J_s/2, J_s/2]$. 
All parameters are expressed in units of the coupling strength $J_s$. 
Depending on the chosen parameter regime, this model displays distinct dynamical behaviors \cite{Pena2021}. Here, we operate in the ergodic region, where information efficiently spreads across the degrees of freedom of the system.

In this work, we consider a quantum input time series
\(\{\rho_k^{(\mathrm{in})}\}_{k=0}^{T-1}\),
where each $\rho_{k}^{(\mathrm{in})}$ represents the density matrix of a $d$-qubit quantum state at time step $k$. The QRC procedure comprises three steps: (i) input information injection to the reservoir, (ii) reservoir evolution, and (iii) extraction of the reservoir features $\mathbf{m}_k$ for computing the output $\mathbf{y}'_k$ through a linear readout \cite{Luko_2009}. 
The system evolution is described by a completely positive trace-preserving (CPTP) map that sequentially incorporates
both input injection $\rho_k^{(\mathrm{in})}$ and the unitary dynamics generated by the Hamiltonian $H$ \cite{Fujii2016} at each time step $k$:
\begin{equation}
    \label{eq:FN map}
    \rho_k = \mathcal{L}_k[\rho_{k-1}] = e^{-iH\Delta t}\big( \rho_k^{(\mathrm{in})} \otimes \mathrm{Tr}_\mathrm{in}\{\rho_{k-1}\} \big)e^{iH\Delta t}\,,
\end{equation}
where $\mathrm{Tr}_\mathrm{in}$ denotes the partial trace over the input subsystem, implementing an effective reset of the input qubits prior to the injection of the new quantum input $\rho_k^{(\mathrm{in})}$.

Observables of the reservoir $\{O_i\}$, here qubit moments (e.g. $\langle\sigma_i^\alpha\rangle, \langle\sigma_i^\alpha\sigma_j^\beta\rangle$), provide the set of computational nodes employed at each time step: $\mathbf{m}_k = \{m_{ik}\}^N_{i=1}$ with $m_{ik}=\langle O_i\rangle_k = \mathrm{Tr}(O_i\rho_k)$. To enhance the expressive power of the reservoir without modifying the linear readout, we employ time multiplexing \cite{Fujii2016, Martínez-Peña2020Information}. This technique samples the unitary evolution at $V$ intermediate times $\frac{v}{V}\Delta t $ (with $1 \leq v \leq V$) between successive input injections, treating each sample as an independent computational node. As a result, the reservoir state vector $\mathbf{m}_k$ is effectively enlarged from dimension $N$ to $NV$.

The readout layer implements a linear regression, $\mathbf{y}'_k = \mathbf{W}_{\mathrm{QRC}}\mathbf{m}_k$, where the state vector $\mathbf{m}_k$ is augmented with a constant bias component (see appendix~\ref{appendix:training_methodology}).  At each time step, the output layer maps the (time-multiplexed) reservoir states to the output values. This formulation reveals a fundamental limitation arising from the structure of quantum reservoir computing. As shown in \cite{Innocenti2023,gili2026learningfunctionsquantumstates}, the map from an input state $\rho_k^{(\mathrm{in})}$ to the expectation values $\langle O_i\rangle_k$ is linear in the input density matrix. Consequently, any observable measured at the output of a quantum reservoir, and thus any linear readout $\mathbf{y}_k' = \mathbf{W}_{\mathrm{QRC}}\mathbf{m}_k$, can only express quantities that are linear functionals of the input states. More formally, for any fixed observable $O$, the map $\rho_k^{(\mathrm{in})} \mapsto \mathrm{Tr}(O\rho_k)$ is linear, meaning that QRC with linear readout cannot compute nonlinear functionals such as polynomials of degree $n>1$ in the density matrix elements. Examples include the purity $\mathrm{Tr}[(\rho_k^{(\mathrm{in})})^2]$ and the von Neumann entropy $-\mathrm{Tr}[\rho_k^{(\mathrm{in})}\log\rho_k^{(\mathrm{in})}]$.

Previous efforts to overcome this limitation have been confined to the QELM framework, where only static inputs are considered \cite{Innocenti2023, gili2026learningfunctionsquantumstates}. Here, we explore an alternative approach based on a hybrid architecture, in which a classical reservoir computing layer introduces additional memory and, more importantly, nonlinear processing capabilities. Our goal is to assess whether this hybrid quantum–classical reservoir can overcome the intrinsic linearity when extracting nonlinear features of past quantum inputs at delayed times $\tau$.

\subsection{Classical RC layer}

The nonlinear behavior is thereby introduced by the classical layer: an ESN consisting of $N_{\mathrm{ESN}}$ internal nodes connected through random weights \cite{Jaeger2001TheechoST, Luko_2009, Yildiz2012, 6105577}. The classical RC procedure is analogous to the one explained before, but in this case the internal reservoir states at time $k$ are represented by the vector $\mathbf{x}_k^{(\mathrm{ESN})}\in\mathbb{R}^{N_{\mathrm{ESN}}}$, whose evolution follows the dynamical map:
\begin{equation}
    \label{eq:ESN_eq}
    \mathbf{x}_{k+1}^{(\mathrm{ESN})} 
    = f\!\left( r\,\mathbf{W}\mathbf{x}_{k}^{(\mathrm{ESN})} 
    + l\,\mathbf{W}^{\mathrm{in}}\mathbf{u}_{k+1} \right),
\end{equation}
where the ESN state is a nonlinear function $f$ of the previous reservoir state and the current input vector $\mathbf{u}_{k+1}\in\mathbb{R}^M$. 
The hyperparameters $r$ and $l$ control the feedback strength and the input gain, respectively, and remain fixed during the training process. The input matrix $\mathbf{W}^{\mathrm{in}}\in\mathbb{R}^{N_{\mathrm{ESN}}\times M}$ mediates the injection of the external signal into the ESN, while the internal reservoir connectivity is encoded in $\mathbf{W}\in\mathbb{R}^{N_{\mathrm{ESN}}\times N_{\mathrm{ESN}}}$.  The entries of both matrices are sampled independently from a uniform distribution on $[-1,1]$ and the internal matrix is rescaled so that its spectral radius is equal to one \cite{Yildiz2012}. In this work, we have selected a logistic function as activation function $f(x)= \frac{1}{1+e^{-x}}$. Finally, as before, the output values are obtained by applying a linear readout, $\mathbf{y}_k = \mathbf{W}_{\mathrm{out}}\mathbf{x}_k^{(\mathrm{ESN})}$. We distinguish between the output values of the quantum and classical reservoir computing schemes, denoted by $\mathbf{y}_k'$ and $\mathbf{y}_k$, respectively.  

\subsection{Hybrid model}

\begin{figure}
  \centering
  \includegraphics[width=\columnwidth]{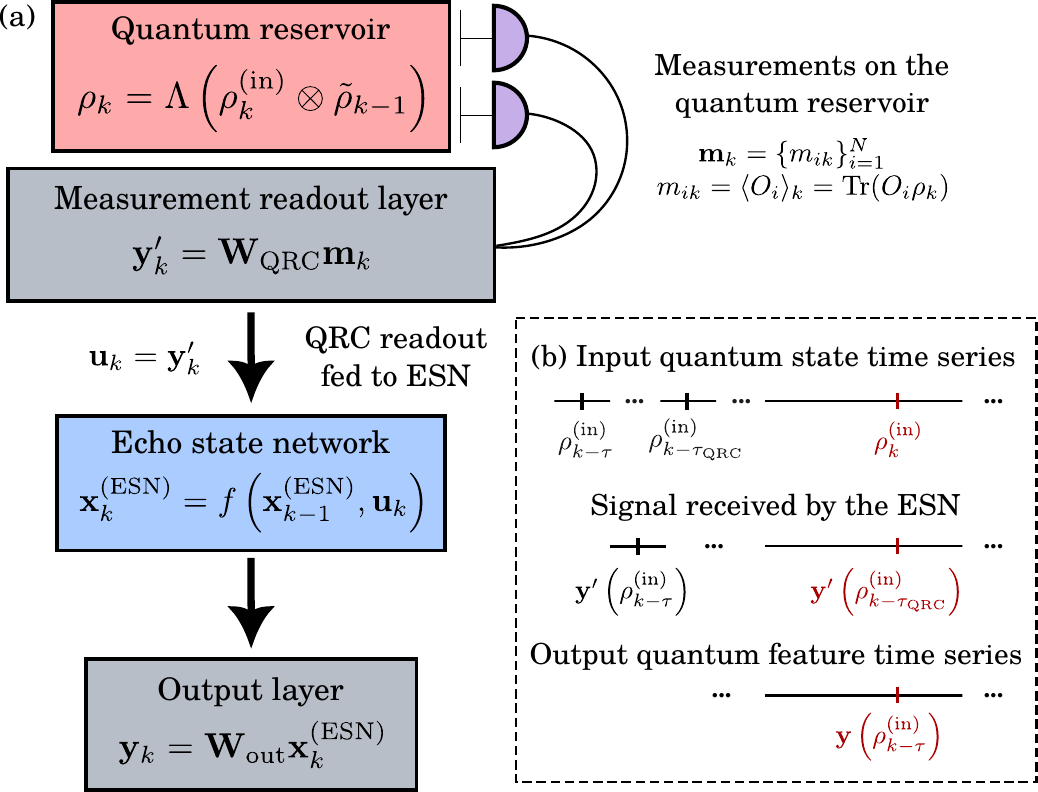}
  \caption{(a) HRC architecture combining a quantum reservoir with a classical ESN. To emphasize the generality of the framework, and since the algorithm is not restricted to the specific maps considered above, we model $\Lambda$ and $f$ as general quantum and classical maps, respectively. (b) Temporal delay: the delayed state target, at a delay $\tau_{\mathrm{QRC}}$, is injected into the classical layer (ESN). This classical vector produced by the QRC readout encodes a tomography-based representation of the quantum input time series and is trained to output features of past input quantum states.  Then the ESN targets non-linear state properties (such as purity or entropy) at a longer delay $\tau\geq\tau_{\mathrm{QRC}}$. The red color indicates that the corresponding values are associated with the same time step in all three sequences.}
  \label{fig:hybrid_diagram}
\end{figure}

The HRC architecture employed in this work is illustrated in figure \ref{fig:hybrid_diagram} and follows the design 
introduced in \cite{Nokkala2024}, where quantum and classical reservoirs jointly process quantum temporal data through their sequential connection. The quantum input is first processed by the QRC. The QRC, whose trained output weights are denoted by $\mathbf{W}_{\mathrm{QRC}}$, acts as a quantum preprocessing stage and is trained to reconstruct the expectation values defining the state tomography of the input at a delay $\tau_{\mathrm{QRC}}$ shorter than the final target delay $\tau$ ($\tau_{\mathrm{QRC}} \leq \tau$). The corresponding target vector is
\begin{equation}
    \label{eq:QRCtarget}
    \mathbf{y}'_k{(\tau_\mathrm{QRC})}
    =
    \bigl(
        \langle O_1\rangle,
        \dots,
        \langle O_{4^{d}-1}\rangle
    \bigr)_{k-\tau_{\mathrm{QRC}}}.
\end{equation}
where $O_i \in \{\mathbb{I}, \sigma^x, \sigma^y, \sigma^z\}^{\otimes d} \setminus \{\mathbb{I}\otimes \mathbb{I}\}$, and $d$ denotes the number of input qubits. 

For a single-qubit input ($d=1$), $\rho^{(\mathrm{in})}:= \rho^{(1)}$, the QRC output vector reduces to the Bloch vector of the input qubit. The corresponding target becomes:
\begin{equation}
\label{eq:QRCtarget_1}
    \mathbf{y}'_k{(\tau_\mathrm{QRC})}
    =
    \bigl(
        \langle \sigma_1^x\rangle,
        \langle \sigma_1^y\rangle,
        \langle \sigma_1^z\rangle
    \bigr)_{k-\tau_{\mathrm{QRC}}}\,. 
\end{equation}
Note that this corresponds to a linear quantum memory task, which the QRC is capable of performing. By enforcing the reconstruction of delayed input states from present reservoir observables, we explicitly exploit the intrinsic memory capacity of quantum reservoirs, demonstrated in previous works at least for classical inputs \cite{Martínez-Peña2020Information}. The reservoir dynamics thereby store and retrieve quantum information before passing it to the subsequent processing stage.

In this framework, the QRC plays the role of an encoder from two complementary perspectives: (i) it transforms quantum data into classical (measurement) data, which is then processed by the ESN, and (ii) by measuring only a limited set of observables\textemdash e.g., the $x$-spin projections of all reservoir qubits\textemdash the QRC is tasked with reconstructing the full tomography of the input state at the earlier delay. This reconstruction exploits the reservoir’s internal dynamics, memory, and information spreading capabilities, without requiring full tomography of the quantum input or measurements along multiple directions.

Once the training of the preprocessing layer has been completed and its parameters are fixed, the classical ESN then receives these reconstructed delayed observables as in equation~\ref{eq:QRCtarget_1}. In other words, the ESN input at time $k$ is $\mathbf{u}_k = \mathbf{y}'_k(\tau_{\mathrm{QRC}})$, and is tasked with predicting nonlinear properties of the original quantum input state at the target delay~$\mathbf{y}_k(\tau)$. In this way, the ESN enhances the nonlinear processing capabilities of the overall architecture, yielding an HRC that can outperform each of its components individually.

\section{Measurement-direction dependence in state reconstruction}
\label{sec: Information Spreading QRC}

Before evaluating the full hybrid architecture, we first analyze the preprocessing stage to understand how the quantum reservoir encodes and retains information about past input states. Specifically, we investigate how measurements of reservoir qubits along a single axis can encode information about all three Bloch components $\left(\langle \sigma_1^x\rangle,\langle \sigma_1^y\rangle,\langle \sigma_1^z\rangle\right)$ of the input qubit at earlier times. This analysis addresses the fundamental question of whether a quantum reservoir, when measured only along one direction, provides sufficient information to reconstruct the full quantum state tomography of past inputs. To quantify the reservoir’s performance in recalling past quantum input tomography, we introduce the so-called capacity, defined as the squared Pearson correlation coefficient between the predicted sequence $\mathbf{y}$ and the target sequence $\bar{\mathbf{y}}$:
\begin{equation}
	\label{eq:capacity}
	C(\tau) = \frac{\mathrm{cov}^2\bigl(\mathbf{y}(\tau), \bar{\mathbf{y}}(\tau)\bigr)}{\sigma^2\bigl(\mathbf{y}(\tau)\bigr)\,\sigma^2\bigl(\bar{\mathbf{y}}(\tau)\bigr)}\,,
\end{equation}
where $\mathrm{cov}(\cdot,\cdot)$ denotes the covariance, and $\sigma^2(\cdot)$ the variance. This metric is bounded in the interval \([0,1]\), with higher values indicating better agreement between the sequences.

\begin{figure}
	\includegraphics[width=\columnwidth]{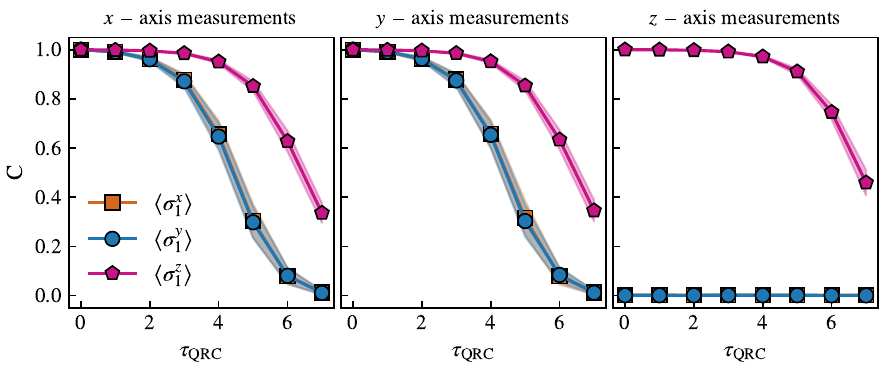}
	\centering
	\caption{Capacity (equation~\ref{eq:capacity}) of the QRC in performing the tomography memory task (equation~\ref{eq:QRCtarget_1}). Each panel corresponds to a different situation, labeled by the plot subtitle. The subtitle $\alpha$-axis measurements indicates that only local observables and two-spin correlations along the $\alpha$ direction of the reservoir are used as computational nodes for linear regression. Results are shown for each of the three axes $\alpha \in \{x,y,z\}$. Each line corresponds to one of the Bloch vector components of the quantum input $\rho_1$ at a given delay $\tau_\mathrm{QRC}$. The plots illustrate the information that a single axis provides about the other projections.}
	\label{fig:QR_prep_xyz_opt}
\end{figure}

Figure \ref{fig:QR_prep_xyz_opt} illustrates this behavior by displaying the system’s capacity to recall the Bloch vector components of qubit~1, with each component represented by a different color, at the optimized parameters found for linear tasks (see table~\ref{tab:hybrid_params}). The analysis is performed separately for the three reservoir-qubit projections $(x,y,z)$, each shown in a different subplot. For a single axis $\alpha$, the local projections $\{\langle \sigma_i^\alpha\rangle\}_i$ and two-spin correlations $\{\langle \sigma_i^\alpha\sigma_j^\alpha \rangle\}_{i,j}$ are considered as computational nodes.

Due to the (anisotropic) dynamics of the reservoir system described by Hamiltonian (equation~\ref{eq:Ising_Hamiltonian}), measurements in different directions are not expected to be equally informative. In particular, it is observed that 
$z$-axis moments of the reservoir qubits do not carry information about the $x$ and $y$ projections of the input\textemdash leading to zero capacity to reconstruct present or previous $\langle \sigma_1^x \rangle $ and $ \langle \sigma_1^y \rangle$ at all delays.  In contrast, measurements along the $x$ and $y$ axes each contain information about all input projections. In all three measurement scenarios, the best reconstructed component is the $z$ projection $\bigl( \langle \sigma_1^z \rangle \bigr)$, as it maintains a capacity close to unity over larger delays than the other projections.  Moreover, we observe that the $x$ and $y$-axis reservoir moments yield virtually identical results for input-state memory. These features reflect the strong influence of the external magnetic field on the reservoir dynamics.

Therefore, measuring only in the $x$ or $y$ direction may suffice to reconstruct faithfully the full input Bloch vector up to $\tau_\mathrm{QRC}=2$, which is crucial since we can avoid measuring the system in non-commutable directions. In the next section, we study how this behavior affects the hybrid architecture's performance.

\begin{table}[t]
\centering
\small
\begin{tabular}{lp{3.5cm}r}
\toprule
Parameters  & Description & Values \\
\midrule
$L$ & Number of qubits & 9 \\
$h$ & Transverse field & 3 \\
$W$ & Disorder strength & 0.01 \\
$\Delta t$ & Unitary evolution time & 5 \\
$V$ & Number of virtual nodes & 10 \\
\midrule
$N_{\mathrm{ESN}}$ & Number of ESN neurons & 45 \\
$r$ & Feedback gain & 4 (linear)\\ & & 0.25 (nonlinear) \\
$l$ & Input gain & $10^{-5}$ (linear)\\ & & $10^{-3}$ (nonlinear) \\
$f$ & Activation function & $f(x) = \frac{1}{1+e^{-x}}$ \\
\midrule
$\tau_\mathrm{QRC}$ & QRC preprocessing target delay & $\lfloor 0.5\,\tau \rfloor$ \\
$T_{\mathrm{wo}}$ & Washout steps & 1000 \\
$T_{\mathrm{train}}$ & Training steps & 2000 \\
$T_{\mathrm{test}}$ & Testing steps & 1000 \\
$N_{\mathrm{iter}}$ & Number of random iterations & 100 \\
\midrule
$\alpha$-axis & QRC observables & $\{\langle \sigma_i^\alpha \rangle\}_{i=1,\dots,L}$\\
& along $\alpha$ & $\cup \{\langle \sigma_i^\alpha \sigma_j^\alpha \rangle\}_{i<j}$ \\
\bottomrule
\end{tabular}
\caption{Simulation parameters of the HRC systems. Distinctions between linear and nonlinear values are made according to the target task. The last row defines the observable set for $\alpha$-axis measurements, including local observables and two-spin correlations, which serve as computational nodes in the QR (both in the hybrid preprocessing layer and in the standalone QRC output layer). Measurements in multiple axes correspond to the union of their respective observable sets. Here, $\lfloor \cdot \rfloor$ denotes the integer part. By default, the memory is shared evenly.}
\label{tab:hybrid_params}
\end{table}

\section{HRC performance}
\label{sec:Hybrid RC Performance}

Once the HRC workflow has been established, the following section benchmarks the system’s performance on both linear and nonlinear tasks from a measurement-restriction perspective and examines how these results scale when using a two-qubit input state by comparing them with those of the individual components. The parameters employed in the simulations are listed in table~\ref{tab:hybrid_params}.

Throughout the rest of this work, reservoir performance is quantified using the normalized mean squared error (NMSE), which emphasizes precision in reproducing target values, rather than correlation, and is defined as:
\begin{equation}
	\mathrm{NMSE}[\mathbf{y},\bar{\mathbf{y}}](\tau) =
    \frac{1}{\sigma^2(\bar{\mathbf{y}})}
    \frac{1}{T}\sum_{k=0}^{T-1} 
    \bigl(y_k(\tau) - \bar{y}_k(\tau)\bigr)^2\,,
\end{equation}
where \(T\) denotes the length of the time series used for evaluation (training or test) and \(y_k\) and \(\bar{y}_k\) are scalar values at time step \(k\). Values \(\mathrm{NMSE}<1\) indicate a prediction error smaller than the variance of the target signal, with lower values corresponding to better reservoir performance. The NMSE may exceed unity when evaluated on testing data.

\subsection{Single-qubit input tasks}

To highlight the role of quantum preprocessing, we evaluate different tasks under various measurement restrictions in the QRC as well as the full-information case. The input sequence consists of quantum states randomly drawn from the Hilbert–Schmidt (HS) ensemble \cite{Zyczkowski_2001}, which is invariant under unitary transformations, implying no preferred measurement direction (see appendix \ref{appendix:Generation_of_random_quanutm_states} for details).

\begin{figure}[b]
	\includegraphics[width=\columnwidth]{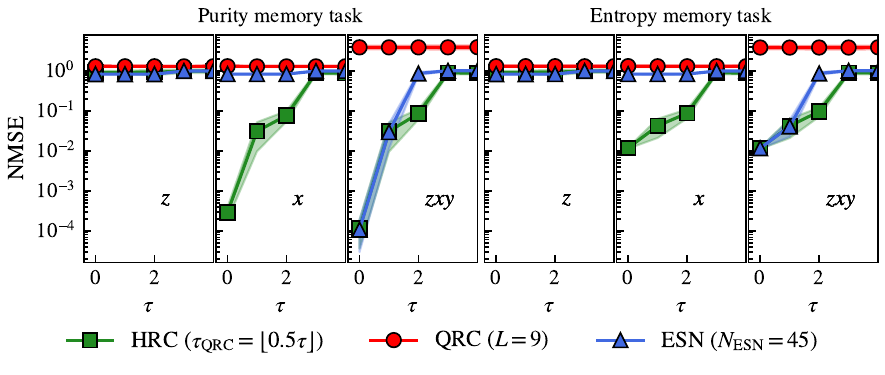}
	\centering
	\caption{Performance of the purity memory task (equation\,\ref{eq:purity_memory_task}, left panel) and the entropy memory task (equation\,\ref{eq:entropy_memory_task}, right panel), for the HRC (green squares), standalone ESN (blue triangles), and standalone QRC (red circles). Each panel corresponds to a different measurement restriction. For example, label $z$ indicates that only local observables and two-spin correlations along the $z$-axis are considered in the QRC and HRC cases, while the standalone ESN receives only the $z$ component of the input state. Shaded regions indicate standard deviations.}
	\label{fig:purity_entropy_hybrid_comparison}
\end{figure}

\textit{Nonlinear tasks}. In figure \ref{fig:purity_entropy_hybrid_comparison} we compare the performance of the standalone QRC (red circles), standalone ESN (blue triangles), and the HRC (green squares) for two different nonlinear memory tasks: (i) the purity memory task (left)
\begin{equation}
	\label{eq:purity_memory_task}
	y_k(\tau) = \mathrm{Tr}\left [\left (\rho^{(\mathrm{in})}_{k-\tau}\right )^2 \right ]\,, 
\end{equation}
and the (ii) entropy memory task (right)
\begin{equation}
	\label{eq:entropy_memory_task}
	y_k(\tau) = S \left ( \rho^{(\mathrm{in})}_{k-\tau}\right ) = - \rho^{(\mathrm{in})}_{k-\tau} \log \left (\rho^{(\mathrm{in})}_{k-\tau} \right )\,.
\end{equation}
For both tasks, we consider the performance with  limited quantum measurements, when these are performed only along the $z$ axis and only along the $x$ axis, and full tomography with Pauli matrices expectation values in all three directions (labeled as the $zxy$ case).

As expected, since both tasks are nonlinear with the quantum input, in figure \ref{fig:purity_entropy_hybrid_comparison} the standalone QRC is unable to perform any of them, leading to an NMSE close to 1 for all delays. Regarding the classical architecture, only after full tomography of the state ($zxy$ case), the ESN achieves a reasonable performance, limited to $\tau=1$. A more interesting situation occurs in the HRC, displaying non-linear memory up to a delay of $2$ even with reduced measurements. As shown in figure \,\ref{fig:QR_prep_xyz_opt}, measuring along the $z$ axis, all architectures are incapable of performing both tasks even at $\tau=0$, since $x$ and $y$ input projections are not accessible.
However, with reduced measurements of $\sigma_x$, the hybrid design is the only one succeeding in the task. Actually, even with reduced access to the injected states, it can overcome the performance of the full-tomography ESN. This demonstrates a clear advantage provided by the preprocessing quantum layer of the HRC with respect to ESN.

Moreover, we can see that even in the full tomography case ($zxy$ panels in figure\,\ref{fig:purity_entropy_hybrid_comparison}), the hybrid setting outperforms the standalone ESN by retaining one more step of memory. This advantage comes from the even delay distribution in the QRC and ESN parts of the hybrid setup for memory tasks\textemdash represented by the parameter $\tau_\mathrm{QRC}$\textemdash when both components have a similar memory (see appendix \ref{appendix:QRCvsESN} for more details). Overall, the hybrid configuration stands out as the best architecture to perform nonlinear tasks, especially in the measurement restriction situation.

Finally, we notice that even if the qualitative behavior is the same in both tasks in figure\,\ref{fig:purity_entropy_hybrid_comparison}, the entropy memory task is more challenging. This is expected as higher-order nonlinearity is needed, leading to reduced performance with respect to the purity memory task.

\textit{Linear task}. To have a more general scope, we extend our analysis to linear tasks, where the standalone QRC is expected to carry out the task. For this purpose, we set the final target of the hybrid setup to match that of the preprocessing layer (equation\,\ref{eq:QRCtarget_1}) but evaluated at a delay $\tau$. This task is hereafter referred to as the state memory task. Then, from the input state projections, we construct the input state $\rho_{k-\tau}^{(\mathrm{in})}$. To evaluate performance we calculate the fidelity \cite{Jozsa01121994} between the input state $\rho^{(\mathrm{in})}$ and the reconstructed state $\hat{\rho}^{(\mathrm{in})}$ obtained from the RC predictions:
\begin{equation}
    \label{eq:fidelity}
    F(\tau) = \left( \mathrm{Tr}\sqrt{\sqrt{\rho_{k-\tau}^{(\mathrm{in})}}\,\hat{\rho}_{k-\tau}^{(\mathrm{in})}\,\sqrt{\rho_{k-\tau}^{(\mathrm{in})}}}\right)^2 \,.
\end{equation}
Then, we use the \textit{infidelity} ($1-F$) to quantify the deviation from perfect reconstruction. The linear readout yields states that are Hermitian with unit trace by construction, but may violate positivity. However, violations occur in a very few samples across all configurations studied. These rare nonphysical instances are excluded from fidelity calculations without affecting the statistical significance of our results.

\begin{figure}
	\includegraphics[width=\columnwidth]{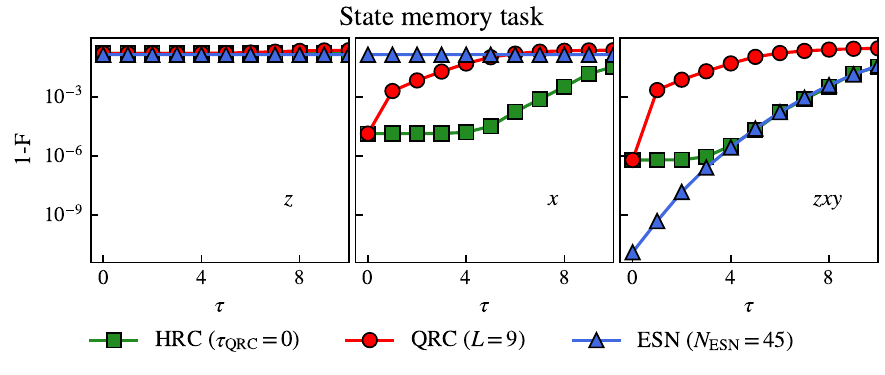}
	\centering
	\caption{Quantum state memory task: Infidelity of state reconstruction at different delays, for the 3 architectures under different measurement schemes as in figure\,\ref{fig:purity_entropy_hybrid_comparison}.}
	\label{fig:fidelity_hybrid_comparison}
\end{figure}

Figure\,\ref{fig:fidelity_hybrid_comparison}, presents the same analysis as for nonlinear tasks. Again, $z$-axis measurements prove insufficient for all three configurations, while $x$-axis measurements can provide the necessary information to reconstruct the state faithfully. Unlike the non-linear tasks presented before, the standalone QRC is now capable of doing the task, although outperformed by the hybrid one that displays a longer memory. The performance of the hybrid design is superior even if the classical substrate alone does not work (vanishing fidelity of the ESN for $x$-axis measurements, central panel in figure\,\ref{fig:fidelity_hybrid_comparison}). Note that in the hybrid configuration we have set a preprocessing delay $\tau_\mathrm{QRC}=0$\textemdash namely, the classical part is in charge of all the memory\textemdash since it provides the best result (see figure\,\ref{fig:effects_of_tau} in the appendix). 

Looking at the full tomography case ($zxy$ panel in figure\,\ref{fig:fidelity_hybrid_comparison}), the standalone ESN achieves the best performance due to the simplicity of the task, which is effectively a short-term memory (STM) task \cite{STM}, as it has access to the full Bloch vector of the input. In contrast, the performance of the HRC is strongly limited by the QRC preprocessing layer. Even when $\tau_{\mathrm{QRC}}=0$, the input to the classical reservoir corresponds to the best estimate provided by the quantum reservoir. Consequently, as shown in the rightmost panel, while the standalone ESN achieves lower infidelity at short delays, the HRC is bounded by the QRC infidelity ($\sim 10^{-6}$ in this case). Similar behavior is observed across different $\tau_{\mathrm{QRC}}$, as shown in figure~\ref{fig:effects_of_tau}.

It is worth assessing whether such precision in state reconstruction leads to meaningful differences. Nevertheless, the hybrid setup retains a clear advantage when full input information is unavailable.

\subsection{Two-qubit input tasks}

We now investigate the HRC capability to process two-qubit states, beginning with generic random states drawn from the HS ensemble (see appendix \ref{appendix:Generation_of_random_quanutm_states}). Note that, in contrast to the single-qubit case, the target dimensionality of the QRC increases from 3 to 15 features, thereby increasing the complexity of the task.

This setting further enables us to address a fundamental property of quantum information, namely, entanglement. To quantify it, we employ the concurrence as a metric, defined as
\begin{equation}
    \label{eq:concurrence}
    \mathcal{C}(\rho) = \max \bigl(0, \Lambda\bigr),\qquad \Lambda = \lambda_1 - \lambda_2 - \lambda_3 - \lambda_4 
\end{equation}
where $\lambda_i$ are the ordered $\bigl(\lambda_1 \geq \lambda_2 \geq \lambda_3 \geq \lambda_4\bigr)$ square roots of the eigenvalues of the Hermitian matrix $R=\sqrt{\sqrt{\rho}\, \tilde{\rho}\,\sqrt{\rho}}$, with the spin-flipped state $\tilde{\rho} = (\sigma^y \otimes \sigma^y) \, \rho^* \, (\sigma^y \otimes \sigma^y)$. The concurrence ranges from $\mathcal{C}=0$ for separable states to $\mathcal{C}=1$ for maximally entangled states, and is valid for all two-qubit density matrices \cite{Wootters1998}.

We consider the same tasks as in the previous sections, that is, purity memory task and state memory task (now for a two-qubit state and partial $x$ measurement, top row), as well as the entanglement memory task (bottom row):
\begin{equation}
\label{eq:entanglement_task}
    y_k(\tau) = \mathcal{C}\left (\rho_{k-\tau}^{(\mathrm{in})}\right)\,.
\end{equation}
To allow the system to learn the concurrence dynamics, we train the model using $\Lambda$ as the target, and apply the maximization of equation \ref{eq:concurrence} only at the testing stage when computing the NMSE. From this point onward, we restrict our analysis to measurements performed along the $x$ direction.

\begin{figure}
	\includegraphics[width=\columnwidth]{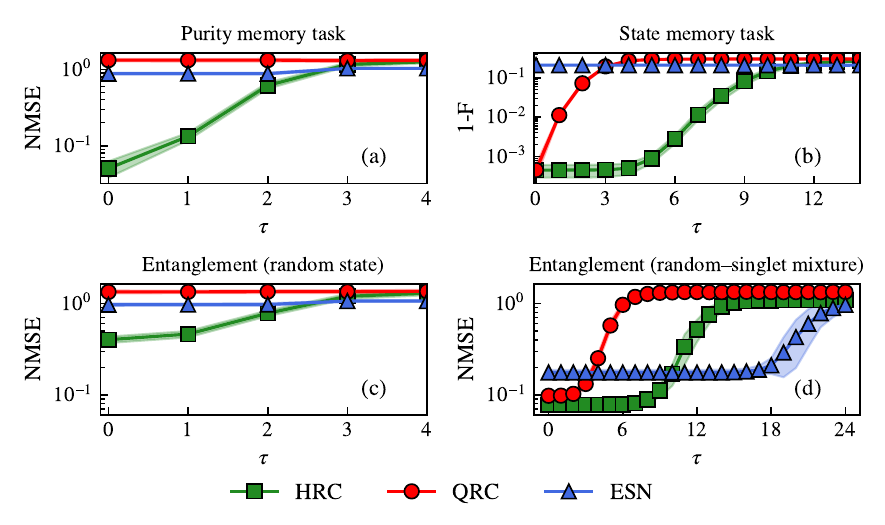}
	\centering
	\caption{HRC performance comparison with a two-qubit input. The top row shows the purity memory task and the state memory task (from left to right) in all settings for partial measurements  $x$. The bottom row presents the entanglement memory task. Panels (a-c) employ generic random two-qubit states, while panel (d) uses a mixture of a random state and a singlet state (equation~\ref{eq:random_werner_state}). Additionally, we employ $\tau_{\mathrm{QRC}} = \tau$ for the left column and $\tau_{\mathrm{QRC}} = 0$ for the right column. For the two-qubit input case, we set $N_{\mathrm{ESN}} = 180$ for both the HRC and the standalone ESN. For the random Werner state, each realization is generated using a random state $\rho_{\mathrm{rand}}$ and a value of the parameter $\alpha$ drawn uniformly from $[0,1]$.}
	\label{fig:two_qubit_performance}
\end{figure}

From the top row of figure\,\ref{fig:two_qubit_performance}, we observe how the hybrid setup still outperforms its standalone components in nonlinear (purity memory, panel (a)) and linear (state memory, panel (b)) tasks. However, the performance at the present step ($\tau=0$) in both situations is hindered by approximately 3 and 2 orders of magnitude, respectively, compared to the $x$-case of the previous section, when a single-qubit input was considered. Moreover, we have noted that in order to retain some memory in the nonlinear task, setting $\tau=\tau_{\mathrm{QRC}}$ is required. This arises from the difficulty of our ESN's configuration in handling nonlinearity, memory and dimensionality reduction of the input at once (see figure~\ref{fig:effects_of_tau} in appendix~\ref{appendix:QRCvsESN}). 

A more challenging scenario is the recall of the input concurrence (equation~\ref{eq:entanglement_task}). We begin by considering completely random states, as done previously. However, we observe that a significant fraction of the generated input states exhibit negative $\Lambda$, which limits the expressivity of the inputs. Moreover, concurrence is a highly nonlinear quantity, and as shown in figure\,\ref{fig:two_qubit_performance} (c), the NMSE in the HRC does not reach sufficiently low values to be considered satisfactory, even though it still outperforms the standalone ESN and QRC.

Consequently, we explore alternative input ensembles to enhance entanglement expressivity. A well-known choice is the family of Werner states \cite{PhysRevA.40.4277}. However, its simplicity allows the concurrence to be directly accessed through $\langle \sigma_1^x \sigma_2^x \rangle$, removing the need for a hybrid approach such as HRC over standalone QRC or ESN. 
For this reason, we consider an intermediate scenario by mixing a random state from the HS ensemble, $\rho_{\mathrm{rand}}$, with the singlet state $\ket{\Psi^-}$, which we refer to as a random–singlet mixture:
\begin{equation}
\label{eq:random_werner_state}
\rho^{(\mathrm{in})} = \alpha\rho_{\mathrm{rand}} + (1-\alpha)\ket{\Psi^-}\bra{\Psi^-}\,.
\end{equation}
This construction provides a broader and more controllable distribution of entanglement, while reducing the overall state complexity depending on the value of $\alpha$ \cite{PhysRevA.111.022412}. We notice that this family of states, when ensemble-averaged, is analogous to a Werner state (see appendix~\ref{appendix:Generation_of_random_quanutm_states} for further details).

Using the random–singlet mixture, figure\,\ref{fig:two_qubit_performance} (d) shows that the task becomes more feasible, with even the standalone QRC able to reconstruct the concurrence at short delays. The HRC provides slightly better performance and extends the memory up to $\tau = 10$. In contrast, the standalone ESN, despite achieving lower overall performance, is able to retain memory for delays approximately twice as long as the HRC.

This behavior may indicate that the entanglement memory task exhibits a degree of quasi-linearity, particularly with respect to the parameter $\alpha$, as illustrated in figure~\ref{fig:conc_dependence_alpha} in appendix~\ref{appendix:Generation_of_random_quanutm_states}. Consequently, additional information from the preprocessing layer provides little improvement at longer delays, unlike in the other scenarios. These observations underscore the importance of the quantum input ensemble for HRC.

\subsection{Nonlinear tasks from state reconstruction}
\label{sec: non-linear tasks from state reconstruction}

While nonlinear tasks are difficult to sustain with high accuracy over long delays, linear task performance remains robust owing to the long memory of the ESN. Given the high fidelity achieved in the state memory task, we introduce an alternative approach to extract nonlinear features from the reconstructed state. Briefly, the procedure involves training the hybrid reservoir to perform the state memory task, followed by a post-processing step external to the RC in which the nonlinear quantity is computed from the reconstructed state. This is conceptually related, though not equivalent, to the idea of next-generation reservoir computing \cite{Gauthier2021Next, PhysRevA.111.022609}, where a nonlinear vector regression is performed.

\begin{figure}
	\includegraphics[width=\columnwidth]{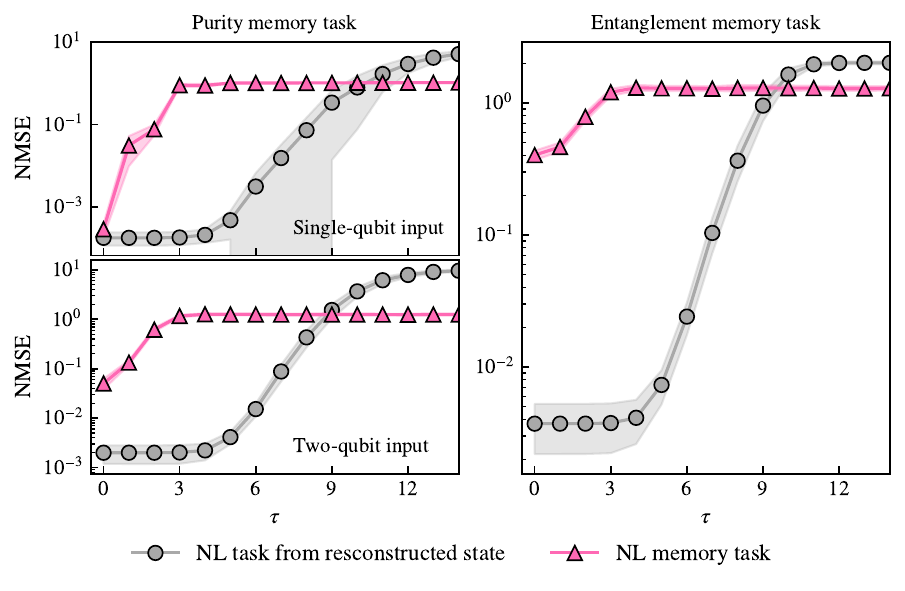}
	\centering
	\caption{Comparison between the performance obtained by targeting the nonlinear (NL) memory task (pink) and by targeting the state memory task followed by computing the nonlinear feature (gray), all with the HRC. The left column shows the purity-memory task for single- and two-qubit cases. The right column corresponds to the concurrence of a general random two-qubit state. Note that the pink lines correspond to the results presented in the previous sections for the respective cases. For the gray lines we set $\tau_{\mathrm{QRC}}=0$.}
	\label{fig:NL_and_Fidelity_comparison.pdf}
\end{figure}

The comparison between targeting the nonlinear task directly (pink triangles) and computing the value from the reconstructed state (gray circles) is depicted in figure\,\ref{fig:NL_and_Fidelity_comparison.pdf}, for the purity memory task (left column) and the 
entanglement memory task (right column). In all cases, this approach shows a significantly better performance. Remarkably, for the entanglement memory task, this alternative method improves the performance by nearly two orders of magnitude compared to the original approach. This result highlights the clear advantage of the proposed method for highly nonlinear features, such as concurrence, for which the original approach struggled.

\section{State perturbation effects}
\label{sec:State Perturbation Effects}

Up to this point, we have considered an ideal measurement process for the quantum system. In this section, we incorporate two important experimental effects affecting QRC performance: quantum back-action on the system induced by measurement, and classical statistical uncertainty arising from a finite ensemble of weak measurements. To do so, we employ the online protocol introduced in \cite{Mujal2023} for time-series processing in QRC architectures. This measurement framework consists of weakly monitoring the system. This is modeled by a weak measurement composed with the unmonitored dynamical map $\mathcal{L}_k$ (equation~\ref{eq:FN map}), or simply after the unitary evolution in the case of time-multiplexing nodes. Such a framework is experimentally relevant, as weak continuous measurements have been demonstrated in several platforms \cite{naghiloo2019introductionexperimentalquantummeasurement,Pan_2020} and can be implemented via partial monitoring \cite{Oriol2026}. The back-action produced by a weak measurement can be modeled by a deterministic map that multiplies the density matrix element-wise:
\begin{equation}
	\rho_k = M \odot \mathcal{L}_k[\rho_{k-1}]\,,
\end{equation}
where $\odot$ denoting the element-wise (Hadamard) product and $M$ is defined as:
\begin{equation}
	M\equiv\tilde{M}^{\otimes L}, \qquad \tilde{M} = \begin{pmatrix}
		1 & e^{-\frac{g^2}{2}} \\
		e^{-\frac{g^2}{2}} & 1 \\
	\end{pmatrix}\,.
\end{equation}

In other words, each time we perform a measurement, the state is updated by a decoherence map that leaves the diagonal elements unchanged while suppressing the coherences (non-diagonal elements) according to the measurement strength $g\geq 0$. In the limit $g=0$, the decoherence map reduces to the unit matrix\textemdash corresponding to the ideal case\textemdash whereas in the limit $g\to\infty$, the matrix $M$ becomes the identity matrix and completely removes the off-diagonal elements of the density matrix.

It is worth noting that increasing the number of virtual nodes reduces the time interval between successive measurements and, consequently, the frequency with which the back-action map is applied. This shortens the unitary evolution time, limiting the system's ability to generate coherences. In the limit of large $g$ and large $V$, the system collapses to one of the eigenstates of the Pauli $z$ operator, as the dynamics effectively correspond to the continuous application of projective measurements \cite{Oriol2026}.

From previous results, we have seen that performing measurements in the $x$ direction is beneficial for our hybrid architecture. Therefore, we employ the following map:
\begin{equation}
	\rho_k = U^\dagger \left( M\odot\left(U\mathcal{L}_k[\rho_{k-1}]U^\dagger\right) \right)U\,,
\end{equation}
where we rotate the system into the $x$ basis using the Hadamard operator
\[U = \tilde{U}^{\otimes L}, \qquad \tilde{U}=\frac{1}{\sqrt{2}} \begin{pmatrix} 1 & 1 \\ 1 & -1 \end{pmatrix}\,,\]
and subsequently rotate it back to the original basis.

\begin{figure}
	\includegraphics[width=\columnwidth]{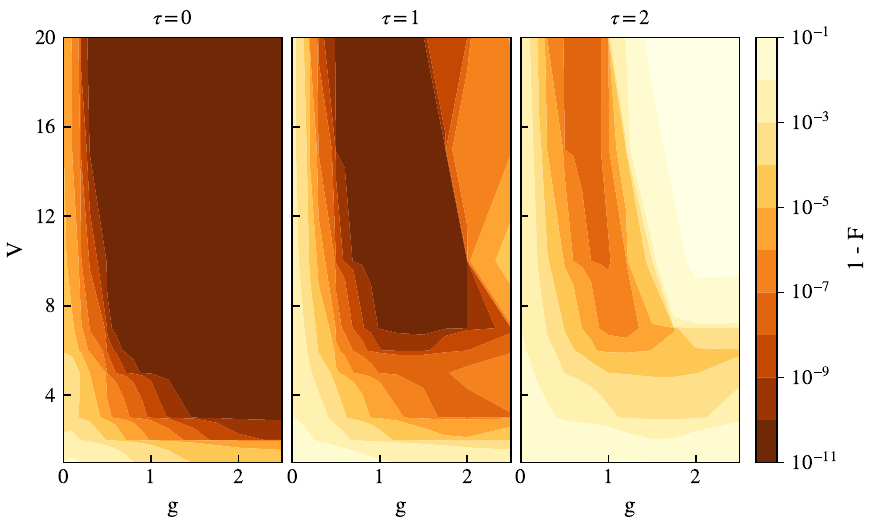}
	\centering
	\caption{State memory task for a single-qubit input as a function of the number of virtual nodes $V$ and the measurement strength $g$ in the quantum system of HRC. The case $g = 0$ corresponds to the ideal measurement scenario. To study the memory dependence on the QRC, we set $\tau_{\mathrm{QRC}} = \tau$, although analyzing the standalone QRC yields similar results. Each point in the heatmap is obtained from 50 random realizations.}
	\label{fig:heatmap_V_vs_g}
\end{figure}

In figure~\ref{fig:heatmap_V_vs_g}, we perform the state memory task by assigning all memory to the quantum layer while varying the number of virtual nodes $V$ and the measurement strength $g$. Interestingly, the back-action not only does not degrade performance, but can even be beneficial in certain parameter regimes, improving upon the ideal case considered previously ($g=0$). This enhancement of QRC performance is consistent with previous studies where dissipation and weak measurements can be optimal for QRC processing of classical time series \cite{Sannia2024dissipationas,  monomi2025feedbackenhancedquantumreservoircomputing,franceschetto}. At zero delay $(\tau=0)$, where no memory is required, measurement back-action appears to assist the system in reconstructing a more accurate state. This advantage is lost for strong measurements as the delay increases. At $\tau = 2$, by contrast, a parameter region still outperforms the ideal case, approximately for $0.2 \lesssim g \lesssim 1.5$ when using 10 virtual nodes.

Despite these encouraging results, an important experimental limitation in the extraction of expectation values has not yet been considered. Following \cite{Mujal2023}, we model the classical statistical uncertainty arising from a finite number of measurement repetitions by adding normally distributed noise to the ideal expectation values, with standard deviations given by
\begin{equation}
	\label{eq:Gaussian_noise}
	s_{\langle \sigma \rangle} = \sqrt{\frac{g^2 + 1}{g^2\,N_{\mathrm{meas}}}}\,,\qquad 	s_{\langle \sigma \otimes \sigma \rangle} = \sqrt{\frac{g^4 + 2g^2 + 1}{g^4\,N_{\mathrm{meas}}}}\,,
\end{equation}
for one and two-qubit observables, respectively. Here, $N_{\mathrm{meas}}$  is the number of measurements per observable per time step.

\begin{figure}
	\includegraphics[width=\columnwidth]{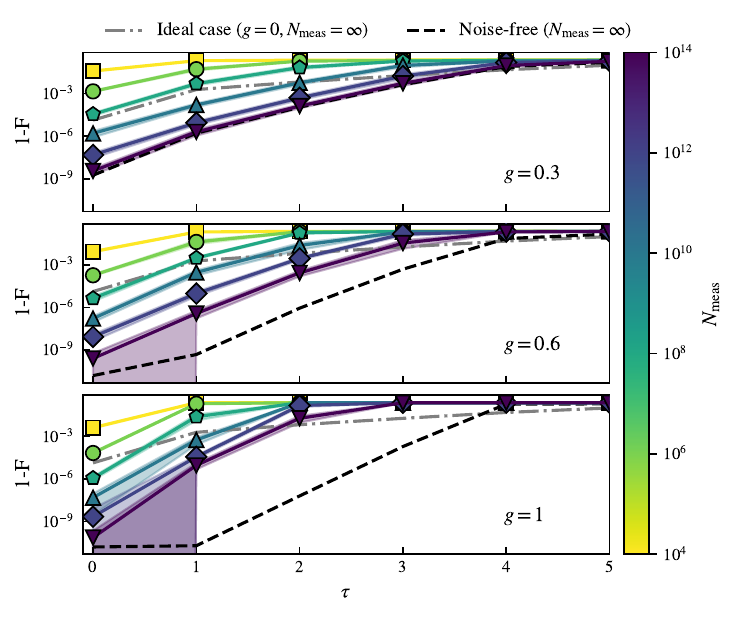}
	\centering
\caption{Effects of statistical noise due to measurements. By setting $g \in \{0.3, 0.6, 1\}$ and $V = 10$, we compare the reservoir's performance when noise (equation\,\ref{eq:Gaussian_noise}) is added to the ideal readouts for $N_{\mathrm{meas}} \in \{10^4, 10^6, 10^8, 10^{10}, 10^{12}, 10^{14}\}$. The plot also shows the noise-free case ($N_{\mathrm{meas}} \to \infty$), denoted by a black dashed line, and the ideal case with neither back-action nor statistical uncertainty ($N_{\mathrm{meas}} \to \infty, g = 0$), represented by the gray point-solid line, for comparison. Each point is obtained from 50 random iterations, each with a different sequence of Gaussian noise and QRC parameters.}
	\label{fig:noise_effects_with_back_action}
\end{figure}

To assess the robustness of the performance under statistical uncertainty, figure~\ref{fig:noise_effects_with_back_action} examines three representative points of the heatmap where measurement back-action outperforms the ideal case ($g \in \{0.3,0.6,1\}$, with $V = 10$), as the number of measurements $N_{\mathrm{meas}}$ is varied. 

As expected, increasing $N_{\mathrm{meas}}$ leads to a reconstructed state that more closely approaches the original one. Statistical noise nonetheless has a significant impact on performance, as evidenced by deviations from the black dashed line, which represents the noise-free results shown in figure~\ref{fig:heatmap_V_vs_g}. When no memory is required ($\tau = 0$), the best performance is achieved for $g = 1$, even in the presence of noise. Using a total number of measurements in the range $N_{\mathrm{meas}} \in [10^{6},10^{8}]$\textemdash with the lower end of this range consistent with current experimental capabilities \cite{Arute2019}\textemdash the system attains performance comparable to that obtained in previous sections (indicated by the gray point–solid line). By contrast, for $N_{\mathrm{meas}}=10^4$ the performance is strongly degraded and the system exhibits no memory in any of the three cases.

When quantum memory is required ($\tau > 0$), statistical effects strongly degrade performance for larger measurement strengths, such as $g = 1$ and $g = 0.6$. In these cases, even unrealistically large numbers of measurements (e.g., $N_{\mathrm{meas}} = 10^{14}$) lead to substantial deviations from the noise-free scenario. By contrast, for $g = 0.3$ the noisy results closely follow the trend of the noise-free case, indicating greater robustness of the memory against statistical uncertainty, although performance with realistic numbers of measurements still falls below the ideal limit.

These results shed light on the experimental feasibility of the online protocol, particularly for the QRC layer, which is typically the most challenging component to implement. When setting $\tau_{\mathrm{QRC}} = 0$, using $g = 1$ and $N_{\mathrm{meas}} = 10^{6}$ yields performance nearly identical to the ideal case, which has already demonstrated excellent results for linear tasks in the hybrid architecture. Conversely, when quantum memory is required, employing weaker measurements (e.g., $g = 0.3$) enables memory performance comparable to the ideal case, at the cost of reduced state reconstruction accuracy in the preprocessing stage, even when using a large number of measurements.

\section{Conclusions}
\label{sec:Conclusions}

We have established HRC as a paradigm for processing quantum time series, directly addressing the fundamental limitation of linearity that constrains pure quantum reservoirs when handling nonlinear functionals of quantum inputs. By sequentially connecting a discrete-variable quantum reservoir, implemented as a disordered spin chain, with a classical echo state network, we have demonstrated that nonlinear tasks such as purity, entropy, and entanglement estimation can be effectively performed while retaining memory of past quantum states.
The superior performance of the hybrid architecture can be understood from two perspectives: (i) the preprocessing of quantum input data into a classical representation, which is subsequently processed by a classical reservoir capable of reliably performing nonlinear tasks, and (ii) the representation of the quantum state in a larger space (information spreading) where the partial tomography can still be useful. Interestingly, the implementation of the HRC system is also robust in realistic scenarios where the system is weakly monitored, allowing continuous processing of temporal information.
The measurement process not only brings the model closer to physical realizations but also enhances performance compared to the idealized case without decoherence \cite{Sannia2024dissipationas, Palacios_2024, franceschetto}. We have further analyzed the system by incorporating noise fluctuations in the measurement outcomes. Despite the degradation in performance, reliable results can still be achieved with current devices.

We find that HRC can outperform its standalone components when measurement-induced information extraction is combined with an appropriate distribution of memory across the quantum and classical reservoirs. When the quantum and classical reservoirs have comparable memory capacities, an even distribution of memory is beneficial. However, if their capacities are highly disparate, allocating memory predominantly to the reservoir with the higher capacity yields better performance. The hybrid architecture outperforms its individual reservoirs in both cases.

Scaling the input dimension to two qubits indicates that relevant quantum features, such as entanglement, can be effectively processed in the hybrid setup, although the choice of input may impact HRC performance, as illustrated by the entanglement memory task. As expected, the hybrid design is likely to exhibit a bottleneck in the embedding of the quantum input, and the HRC may suffer from limitations similar to those encountered in classical machine learning.

Although we have addressed the main challenge of processing nonlinear quantum tasks, our results show that the classical component may still struggle with nonlinearity, dimensionality reduction—from input to output—and memory requirements. To mitigate these limitations, we proposed targeting linear state reconstruction followed by post-processed nonlinear computation. This approach has proven effective, improving entanglement estimation by nearly two orders of magnitude compared to direct nonlinear targeting.
Other recent approaches, such as feedback protocols \cite{Kobayashi_2024} combined with weak measurements \cite{monomi2025feedbackenhancedquantumreservoircomputing}, may offer additional benefits for quantum time-series processing and are promising avenues for future exploration. Moreover, the effects of noisy classical components and their impact on the overall hybrid system merit further study. Extending the present framework to purely classical time series is also a potential direction.

Finally, HRC provides a practical, near-term route for quantum machine learning on quantum temporal data. The architecture is compatible with existing qubit platforms (superconducting qubits \cite{PhysRevLett.120.050507,Suzuki2022}, trapped ions \cite{Zhang2017}, optical lattices \cite{Choi_2016}) and robust under weak measurements and finite sampling. We expect hybrid designs to become a standard tool for processing quantum state sequences in noisy, realistic settings.

%--------------------------------------------
\begin{acknowledgments}
We acknowledge MINECO through the QUANTUM SPAIN project, and EU through the RTRP - NextGenerationEU within the framework of the Digital Spain 2025 Agenda. We also acknowledge the Spanish State Research Agency, through the COQUSY project PID2022-140506NB-C21 and -C22 funded by MICIU/AEI/10.13039/50110001103 and by ERDF, EU; through the QuantCom project CNS2024-154720, funded by MICIU/AEI/10.13039/501100011033 and cofunded by the European Union; the Quantera II program that has received funding from the EU's H2020 research and innovation program under Grant Agreement No. 101017733, and from the Spanish State Research Agency  (project Coquadis PCI2024-153446) funded by MICIU/ AEI/10.13039/50110001103; and through the María de Maeztu project CEX2021-001164-M, funded by MICIU/AEI/10.13039/501100011033.
The CSIC Interdisciplinary Thematic Platform (PTI+) on Quantum Technologies in Spain (QTEP+) is also acknowledged.
\end{acknowledgments}

% \funding{Sample text inserted for demonstration.}
% % This section is a list of funder names and grant numbers

% \roles{Sample text inserted for demonstration.}
% % List author names and the contributions made to the article, using terms from the NISO Contributor Roles Taxonomy (CRediT) https://credit.niso.org

\section*{Conflicts of Interest}
The authors declare no conflicts of interest.
\section*{Data Availability}
Data is available from the corresponding author at reasonable request.

% -------- Bibliography --------
\bibliography{references.bib}
\appendix
\onecolumngrid

\section{Generation of random quantum states}
\label{appendix:Generation_of_random_quanutm_states}

The generation of a diverse set of quantum input states plays a crucial role in benchmarking reservoir performance. Several methods exist for the numerical generation of random quantum states \cite{Maziero2015}. In this work, we focus on generating random mixed states by sampling density matrices from the Hilbert–Schmidt ensemble. The procedure is as follows \cite{Zyczkowski_2001}:
\begin{enumerate}
    \item Generate a random $2^L \times 2^L$ complex matrix $M$ whose entries are
    independently drawn from a normal distribution with zero mean and unit variance,
    i.e.\ $M_{ij} = \mathcal{N}(0,1) + i\,\mathcal{N}(0,1)$.
    \item Construct the density matrix
    \begin{equation}
		\label{eq:random_qubit_generator}
        \rho = \frac{MM^\dagger}{\mathrm{Tr}(MM^\dagger)}\,,
    \end{equation}
    which is Hermitian, positive semidefinite, and has unit trace.
\end{enumerate}
This construction corresponds to sampling $\rho$ from the Hilbert–Schmidt ensemble, which is invariant under unitary transformations.
\subsection{Random-singlet mixture}

\begin{figure}
	\includegraphics[width=0.55\textwidth]{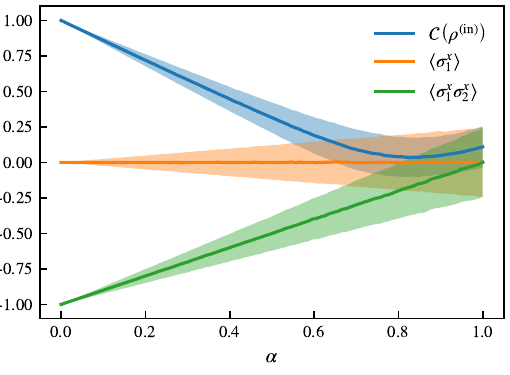}
	\centering
	\caption{Concurrence and $x$-moments of the biased entangled state (equation~\ref{eq:random_werner_state}) as a function of the mixture parameter $\alpha$. For each point, $10^4$ random states $\rho_{\mathrm{rand}}$ are generated according to the HS measure. The value of $\langle \sigma_2 ^x \rangle$ is omitted since it is equivalent to $\langle\sigma_1^x \rangle$.}
	\label{fig:conc_dependence_alpha}
\end{figure}

With the aim of obtaining higher entangled states\textemdash hence, higher values of concurrence\textemdash in equation\,\ref{eq:random_werner_state} we propose a biased entangled random state, in which the ensemble behaves as a Werner state:
\begin{equation}
	\mathbb{E}[\rho] = \alpha \mathbb{E}[\rho_{\mathrm{rand}}] + (1-\alpha)\ket{\Psi^-}\bra{\Psi^-} = \frac{\alpha}{4}\mathbb{I} + (1-\alpha)\ket{\Psi^-}\bra{\Psi^-} \equiv \rho_W\,.
\end{equation}
As explained in the main text, it is composed of a mixture of a random state from the Hilbert-Schmidt ensemble (equation\,\ref{eq:random_qubit_generator}) and the singlet state $\ket{\Psi^-}$. As we have remarked, the concurrence appears to be linear with the state since QRC is able to target the values. Thus, in figure\,\ref{fig:conc_dependence_alpha} we plot the value of the Concurrence and two moments of the input state against the mixture parameter $\alpha$ to see their dependence. It is worth highlighting how the closer the state is to the antisymmetric Bell state\textemdash namely, $\alpha<0.5$ \textemdash the clearer is the linear dependence on $\langle \sigma_1^x \sigma_2^x \rangle$ and all information is stored in this quantity. This explains why the ESN, when fed only the $x$ moments as inputs, is able to retain memory, as shown in figure~\ref{fig:two_qubit_performance}. On the other hand, the more we approach the random state, this unique dependence is lost, and other moments contribute, leading to a more complex nonlinear value.

\section{Training methodology}
\label{appendix:training_methodology}

For the training procedure of the output layer, the time sequence is divided into three segments, such that $T = T_{\mathrm{wo}} + T_{\mathrm{train}} + T_{\mathrm{test}}$. The first $T_{\mathrm{wo}}$ steps are used to wash out the influence of the initial conditions, and are discarded from further processing. For the remaining sequence, linear regression (LR) is applied to obtain the predicted values:
\begin{equation}
    \mathbf{y} = \mathbf{W}^{\mathrm{out}} \mathbf{x}\,,
\end{equation}
where $\mathbf{x}$ and $\mathbf{y}$ are matrices of dimensions $N \times T_{\mathrm{train}}$ and $M \times T_{\mathrm{train}}$, respectively. Consequently, $\mathbf{W}^{\mathrm{out}}$ is a matrix of dimensions $M \times N$. Note that $M$ and $N$ are the dimensions of the subsets of the output signal and reservoir states, respectively. Moreover, we include a bias term in the reservoir states, which is equivalent to adding a row with $\mathbf{x}_k = 1$ for all $k$. In this case, the dimension $N$ increases to $N+1$.

Since RC is a supervised algorithm, the system is trained according to the target values $\mathbf{\bar{y}}$, which represent the true outputs associated with the input signals. For this purpose, the target values are split into two time series corresponding to $T_{\mathrm{train}}$ and $T_{\mathrm{test}}$, respectively, and the same partition is applied to $\mathbf{x}$. In LR, the goal is to determine the elements of the matrix $\mathbf{W}^{\mathrm{out}}$ such that the mean squared error (MSE) between the predicted and target values is minimized:
\begin{equation}
\label{eq:MSE}
    \mathrm{MSE}[\mathbf{y}, \bar{\mathbf{y}}]{(T_{\mathrm{train}})} = \frac{1}{T_{\mathrm{train}}}\sum_{k=0}^{T_{\mathrm{train}}-1} \|\mathbf{y}_k - \bar{\mathbf{y}}_k\|^2\,,
\end{equation}
where $\mathbf{y}_k$ denotes the $k$-th column of the matrix $\mathbf{y}$ (predicted outputs), and $\bar{\mathbf{y}}_k$ denotes the corresponding target column. To achieve this, we apply the Ordinary Least Squares (OLS) method to obtain the estimator matrix:
\begin{equation}
\label{eq:OLS method}
    \widehat{\mathbf{W}}^{\mathrm{out}} = \mathbf{\bar{y}} \mathbf{x}^T (\mathbf{x}\mathbf{x}^T)^{-1} = \mathbf{\bar{y}} \mathbf{x}^+\,,
\end{equation}
where $\mathbf{x}^+ = \mathbf{x}^T (\mathbf{x}\mathbf{x}^T)^{-1}$ denotes the Moore--Penrose pseudo-inverse \cite{moore1920reciprocal,penrose1955generalized}. Although computing the pseudo-inverse can be computationally expensive, in practice we rely on the Python implementation provided in \texttt{Scikit-learn} \cite{scikit-learn}. Other approaches, such as Ridge regression, introduce a regularization parameter to reduce the risk of overfitting. However, in our case, we restrict ourselves to LR due to its simplicity and because we consider sufficiently long time series.

Once LR is completed, the $\mathbf{x}$ values from the test window are used to compute the predicted outputs:
\begin{equation}
    \mathbf{y}^{(\mathrm{pred})} = \widehat{\mathbf{W}}^{\mathrm{out}}\mathbf{x}^{(\mathrm{test})}\,.
\end{equation}

\section{ESN characterization}

\begin{figure}
    \centering
    \includegraphics[width=0.7\textwidth]{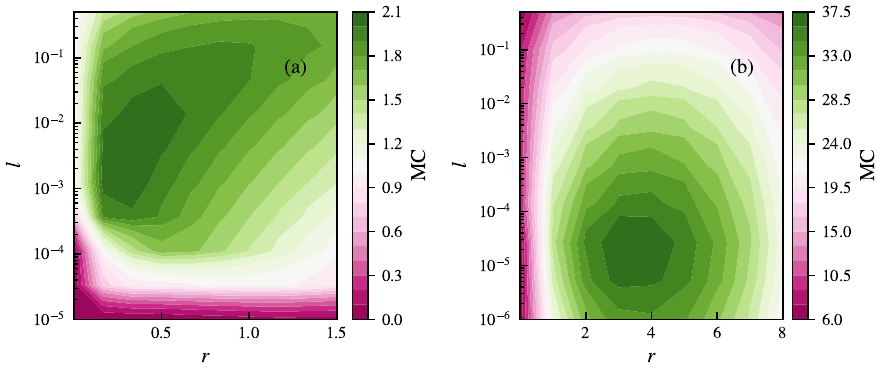}
    \caption{Total memory capacity (equation~\ref{eq:memory_capacity} for (a), equation~\ref{eq:memory_capacity_vector} for (b)) for various combinations of the feedback gain $r$ and input gain $l$. The parameter $l$ is plotted on a logarithmic scale, while $r$ is linear. (a) Purity memory task with (equation~\ref{eq:purity_memory_task}) with $\tau_{\mathrm{max}}=10$. (b) State memory task (equation~\ref{eq:QRCtarget_1}) with $\tau_{\mathrm{max}}=50$. Each point in the map is the mean of 100 random configurations.}
    \label{fig:esn_opt}
\end{figure}

The optimal classical hyperparameters $r,\,l$ (equation\,\ref{eq:ESN_eq}) in the logistic function, reported in table~\ref{tab:hybrid_params}, were selected according to the results shown in figure~\ref{fig:esn_opt}. The total memory capacity
\begin{equation}
\label{eq:memory_capacity}
MC = \sum_{\tau} C(\tau)
\end{equation}
was employed as a figure of merit to quantify the optimal values. Theoretically, the maximum bound of this sum is infinite. Computationally, we set a sufficiently large delay such that $C(\tau_{max})\sim 0$.

As a reference, to probe nonlinearity we use the purity memory task (equation~\ref{eq:purity_memory_task}), while for linear performance we use the state memory task (equation~\ref{eq:QRCtarget_1}). For the latter, since we obtain a capacity value $C_\alpha$ for each delayed Bloch vector component $y_{k}^{(\alpha)}(\tau) = \langle \sigma_1^\alpha\rangle_{k-\tau}$, we redefine the total memory capacity as the sum over all components: 
\begin{equation}
\label{eq:memory_capacity_vector}
MC=\sum_{\alpha}\sum_\tau\,C_\alpha(\tau)\,, 
\end{equation}
with $\alpha \in \{x,y,z\}$.

Although this is not the primary focus of the present work, we note that the input gain is much smaller than the feedback, particularly for linear tasks. Nevertheless, these values are considered optimal based on the results shown in the plots. Similar behavior is observed for other linear and nonlinear tasks, so we simplify our analysis by focusing on these two representative cases.
\section{QRC vs ESN}
\label{appendix:QRCvsESN}

Building on the previous discussion, we now compare QRC and ESN components in the hybrid setting from two perspectives: the delay distribution, represented by $\tau_\mathrm{QRC}$, and the dimensionalities of their respective reservoirs, $L$ and $N_{\mathrm{ESN}}$.

\subsection{Delay distribution between classical and quantum reservoirs}

We restrict our analysis to three memory-distribution cases: (i) $\tau_\mathrm{QRC}=\tau$, where all the memory is carried by the quantum reservoir, (ii) $\tau_\mathrm{QRC}=0$, where memory lies entirely in the ESN, and (iii) $\tau_\mathrm{QRC}=\lfloor \tau/2 \rfloor$ (with $\lfloor\cdot\rfloor$ for the integer part), where memory is shared evenly.

\begin{figure}
	\includegraphics[width=0.7\textwidth]{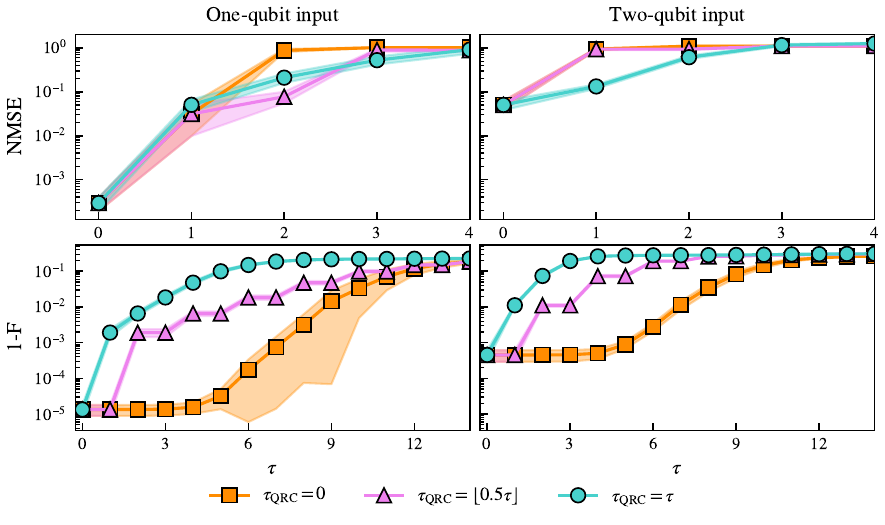}
	\centering
    \caption{HRC performance for different values of $\tau_\mathrm{QRC}$ for the one-qubit (left) and two-qubit (right) cases. The first row corresponds to the purity memory task, while the second row represents the state memory task. Notice that for $\tau=0,1$, $\tau_\mathrm{QRC}=0$ and $\tau_\mathrm{QRC}=\lfloor\tau/2\rfloor$ coincide.}
    \label{fig:effects_of_tau}
\end{figure}

In figure\,\ref{fig:effects_of_tau} we compare these cases for a nonlinear task\textemdash where the first row shows the purity memory task for both one and two qubit input\textemdash and for a linear task\textemdash where the bottom row represents the state memory task for both inputs. As expressed in the main text, for linear tasks the best situation is provided by letting the classical part carry all the weight of memory, setting $\tau_\mathrm{QRC}=0$. On the other hand, the ESN memory for nonlinear tasks is notoriously hindered and hence for the one-qubit case and even distribution provides the best result, whereas for a two-qubit input the quantum part handles the memory of the hybrid setup.

\subsection{Reservoir size}

We now study the relative importance of the quantum and classical components of the HRC architecture by varying their respective sizes (and thus the number of computational nodes). Recall that increasing the number of ESN nodes (denoted by $N_{\mathrm{ESN}}$) modifies the size of the output layer $\mathbf{W}_{\mathrm{out}}$, while enlarging the quantum reservoir (denoted by $L$) affects the preprocessing layer $\mathbf{W}_{\mathrm{QRC}}$. Both changes influence performance, but in different ways.

\begin{figure}
	\includegraphics[width=\textwidth]{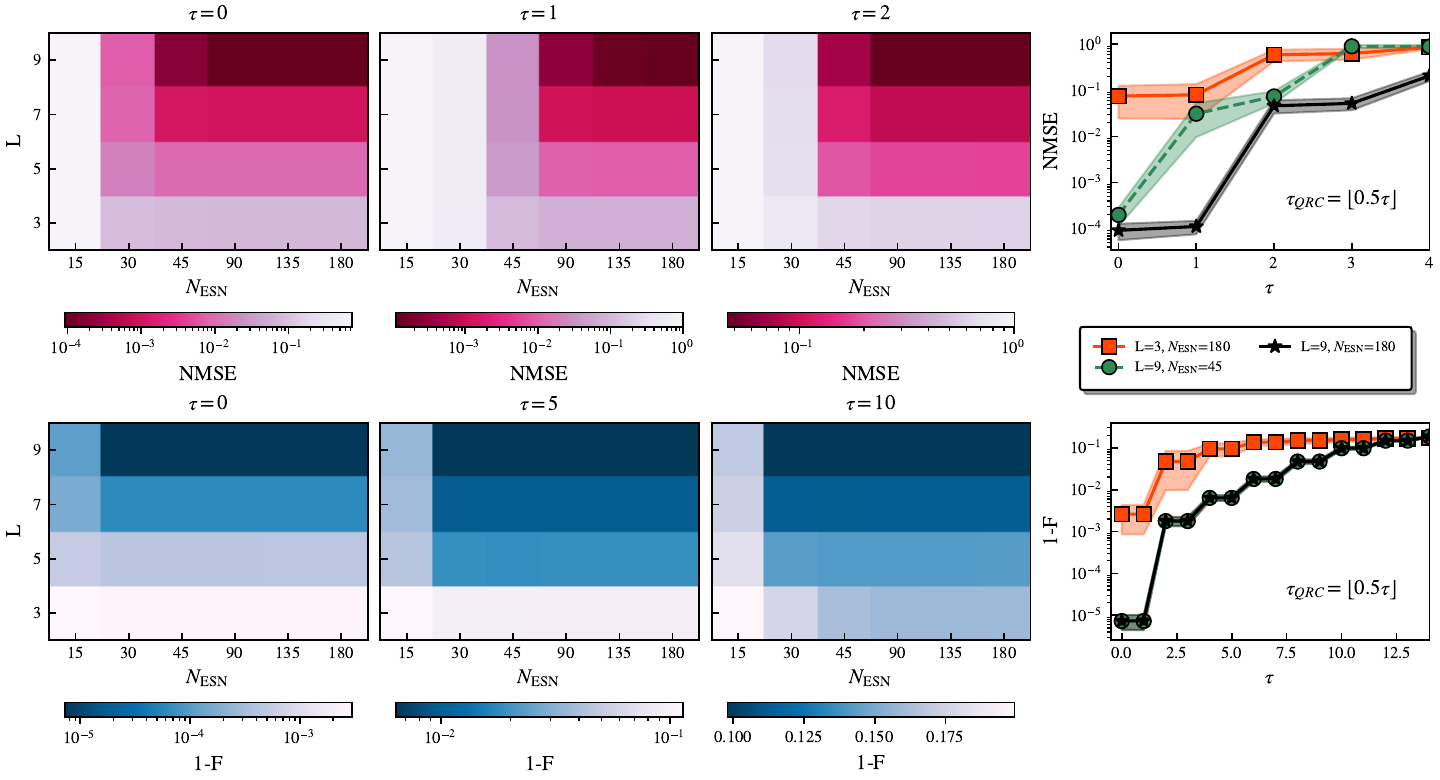}
	\centering
	\caption{Performance of the hybrid architecture for the purity memory task (top row) and state memory task (bottom row) for various combinations of quantum and classical reservoir sizes. The reservoirs are determined by the number of qubits $L$ in QRC and the number of neurons $N_{\mathrm{ESN}}$ in the ESN. In both tasks only $x$-axis measurements and $\tau_\mathrm{QRC}=\lfloor 0.5 \tau \rfloor$ are considered. The first three columns show the variation in performance for $\tau\in\{0,1,2\}$ in the purity memory tasks and $\tau\in\{0,5,10\}$ for the state memory task. The last column displays the performance as a function of $\tau$ for selected combinations $(L,N_{\mathrm{ESN}}) \in \{(3,180), (9,45), (9,180)\}$. The green dashed line is included as a reference, corresponding to the setup considered in the main text.}
    \label{fig:qrc_vs_esn}
\end{figure}

Figure~\ref{fig:qrc_vs_esn} shows two-dimensional plots of how performance in the purity memory task (top row) and the state memory task (bottom row) depends on the number of qubits $L$ and classical neurons $N_{\mathrm{ESN}}$ at different delays. We consider an even distribution of delays to assess the potential and limitations of both components.

In general, as expected, larger reservoirs in both components yield better performance. For the nonlinear task, a small ESN (around 15 nodes) fails to perform adequately regardless of the QR size, while even a small QRC ($L=3$) with a larger ESN can achieve NMSE $\sim 10^{-1}$ at short delays. Intuitively, larger delays require larger dimensions in both reservoirs. Interestingly, increasing the number of classical nodes beyond 90 or 135 yields little additional improvement, whereas enlarging the quantum reservoir continues to enhance performance, provided the ESN is sufficiently large. This highlights a potential limitation since increasing the ESN size is technically less demanding than scaling the number of qubits. Nevertheless, we can observe in the rightmost panel, with $L=9$ and $N_{\mathrm{ESN}}=180$ (black line), the error remains relatively small even at $\tau=4$.

For the linear task, we again observe the constraint imposed by our delay-distribution configuration, which is advantageous only when both the quantum and classical reservoirs have comparable memory. As shown in the bottom row of figure~\ref{fig:qrc_vs_esn}, when the quantum memory is shorter than the classical memory, performance becomes insensitive to increasing the ESN size, as illustrated by the overlap of the green and black lines in the bottom-right panel.

\end{document}